# Giant effective damping of octupole oscillation in an antiferromagnetic Weyl semimetal

*Shinji Miwa[1,2,3]\*, Satoshi Iihama[4,5], Takuya Nomoto[3,6]\*\*, Takahiro Tomita[1,3], Tomoya Higo[1,3], Muhammad Ikhlas[1], Shoya Sakamoto[1], YoshiChika Otani[1,2,3,7], Shigemi Mizukami[4,5,8], Ryotaro Arita[3,6,7], and Satoru Nakatsuji[1,2,3,9]*

[1]The Institute for Solid State Physics, The University of Tokyo, Kashiwa, Chiba 277-8581, Japan

[2]Trans-scale Quantum Science Institute, The University of Tokyo, Bunkyo, Tokyo 113-0033, Japan

[3]CREST, Japan Science and Technology Agency (JST), Kawaguchi, Saitama 332-0012, Japan

[4]Advanced Institute for Materials Research, Tohoku University, Sendai, Miyagi 980-8577, Japan

[5]Center for Spintronics Research Network (CSRN), Tohoku University, Sendai, Miyagi 980-8577, Japan

[6]Department of Applied Physics, The University of Tokyo, Tokyo 113-8656, Japan

[7]RIKEN, Center for Emergent Matter Science (CEMS), Saitama 351-0198, Japan

[8]Center for Science and Innovation in Spintronics (CSIS), Tohoku University, Sendai, Miyagi 980-8577, Japan

[9]Department of Physics, The University of Tokyo, Tokyo 113-0033, Japan

\*miwa@issp.u-tokyo.ac.jp

\*\*nomoto@ap.t.u-tokyo.ac.jp.

ABSTRACT:

A magnetic Weyl semimetal is a recent focus of extensive research as it may exhibit large and robust transport phenomena associated with topologically protected Weyl points in momentum space. Since a magnetic texture provides a handle for the configuration of the Weyl points and its transport response, understanding of magnetic dynamics should form a basis of future control of a topological magnet. $Mn_3Sn$ is an example of an antiferromagnetic Weyl semimetal that exhibits a large response comparable to the one observed in ferromagnets despite a vanishingly small magnetization. The non-collinear spin order in $Mn_3Sn$ can be viewed as a ferroic order of cluster magnetic octupole and breaks the time-reversal symmetry, stabilizing Weyl points and the significantly enhanced Berry curvature near the Fermi energy. Here we report our first observation of time-resolved octupole oscillation in $Mn_3Sn$. In particular, we find the giant effective damping of the octupole dynamics, and it is feasible to conduct an ultrafast switching at < 10 ps, a hundred times faster than the case of spin-magnetization in a ferromagnet. Moreover, high domain wall velocity over 10 km/s is theoretically predicted. Our work paves the path towards realizing ultrafast electronic devices using the topological antiferromagnet.

TEXT:

A Weyl semimetal is a topological system in three dimensions, which appears in either time-reversal-symmetry (TRS) or inversion-symmetry broken state [1-4]. It is characterized by gapless electronic excitations called Weyl fermions, formed as a result of a linear crossing of two non-degenerate bands at a pair of momentum points with different chiralities. As these Weyl points can be viewed as unit-size monopole of underlying Berry curvature, Weyl semimetals may exhibit various exotic phenomena such as Fermi arc, anomalous Hall and Nernst effects (AHE and ANE), chiral anomaly, and gyrotropic effects [1-18]. In particular, for TRS broken or magnetic Weyl semimetals, magnetic order determines the configuration of the Weyl points in momentum space and provides the handle for controlling the transport responses [10,12-18]. Therefore, understanding of the magnetic dynamics forms the basis for future research and application of Weyl semimetals.

In the field of spintronics, antiferromagnetic (AF) metals have attracted significant attention as next-generation active materials of electronic devices for their vanishingly small stray field perturbing neighboring cells. The recent rapid development in AF spintronics [19-21] has led to the demonstration of electric reading and writing of an AF state [22,23], which has been recently further supported by several kinds of AF domain imaging techniques [24-28]. Because of its vanishingly small magnetization, the detection means for such an AF metallic state has been restricted to anisotropic magnetoconductance [22,23], quadratic magneto-optical effects [29], and resonant X-ray diffraction [30], which are far weaker than the magnetization $M$-linear response such as AHE and magneto-optical effects employed for ferromagnets.

With this respect, recently discovered $D0_{19}$-Mn$_3$Sn stands out as a unique antiferromagnet that exhibits large electric and optical $M$-linear responses such as AHE [10], ANE [13,14], and magneto-optical Kerr effect (MOKE) [31] even though it has only a vanishingly small magnetization. Significantly, it has been clarified that Mn$_3$Sn is a magnetic Weyl semimetal [10,13,15]. The non-collinear chiral magnetic texture in Mn$_3$Sn can be viewed as a ferroic order of a cluster magnetic octupole and breaks TRS macroscopically [32]. Figure 1a shows the crystal and magnetic structures of Mn$_3$Sn. The magnetic moments of Mn lie in the (0001) plane and form an inverse triangular spin structure. In this structure, one unit of the cluster magnetic octupole is made of the six neighboring moments on an octahedron (bi-layered triangle) of Mn atoms as featured by a colored hexagon in Figs. 1a and 1b. By 180° rotation of each spin, the octupole polarization may reverse its direction (Fig. 1b). As a result, the in-plane non-collinear spin order can be viewed as Q = 0 order of the magnetic octupole. The AF spin texture with this ferroic order induces large Berry curvature due to Weyl points in momentum space [13,15], leading to the large transverse response. In this regard, the observation of the spin dynamics in a topological magnet would be a key step for future manipulation of the Weyl points in the momentum space and the associated large responses. In addition, similarly to the case of the spin-magnetization dynamics in ferromagnetic metals, understanding and control of the time-dependence of the cluster magnetic octupole in a chiral AF metal would be an important step for developing the device physics in magnetism. To date, there has been no report on the time-resolved observation of spin dynamics in either ferromagnetic or antiferromagnetic Weyl semimetals. Even when we focus on the previous research on antiferromagnets [29,30,33-37], the direct, time-resolved observation of the spin precession has been limited to AF insulators [35-37] and never been made in AF metals. A few papers report the time-resolved dynamics of the order parameter in AF metals [29,30,34] but never been able to associate them with the spin-wave modes. In this article, we report our observation of the time-resolved spin dynamics in the AF Weyl semimetal $Mn_3Sn$.

Let us first discuss the case of collective spins in ferromagnetic materials, where each spin is coupled in parallel by exchange interaction (Supporting Information 1.). A precessional motion of each spin is always in-phase as schematically shown in Fig. 1c, and thus the energy scale of the resonant frequency ($\hbar\omega \sim K$) is independent of the exchange interaction and is determined by $K$ only, where $K$ is magnetic anisotropy energy mainly originating from on-site spin-orbit interaction. A typical time scale for the magnetization switching, expressed as $(\Delta\omega)^{-1}$, can be $\sim(2\alpha K)^{-1}\hbar$. Here $\Delta\omega$ and $\alpha$ are spectral linewidth and effective damping constant, respectively. Generally, the time scale is longer than 1 ns (e.g. $\omega$ = 10 GHz and $\alpha$ = 0.1).

To consider collective spins in a chiral AF metal (Supporting Information 1.), the following Hamiltonian to treat the inverse triangular spin structure can be employed [38],

$$\mathcal{H} = J\sum_{\langle ia,jb\rangle} \mathbf{S}_{ia}\cdot\mathbf{S}_{jb} + D\sum_{\langle ia,jb\rangle}\varepsilon_{ab}\mathbf{z}\cdot\left(\mathbf{S}_{ia}\times\mathbf{S}_{jb}\right) - \frac{K}{2}\sum_{ia}\left(\mathbf{k}_a\cdot\mathbf{S}_{ia}\right)^2. \quad (1)$$

Here, **S**, $J$ and $D$ denote spin-angular momentum, exchange interaction and Dzyaloshinskii-Moriya interaction, respectively. ($i, j$) and ($a, b$) refer to the Mn sites and one of the three sublattices ($A$, $B$, and $C$) of the inverse triangular lattice structure, respectively. $K$ is introduced to describe six-fold magnetic anisotropy in Mn$_3$Sn with $\mathbf{k}_a$ = (cos$\psi_a$, sin$\psi_a$, 0) and ($\psi_A$, $\psi_B$, $\psi_C$) = (0, 4π/3, 2π/3). $\varepsilon_{ab}$ is the antisymmetric tensor which satisfies $\varepsilon_{AB} = \varepsilon_{BC} = \varepsilon_{CA} = 1$ and **z** is the unit vector along the c-axis. Note that in-plane and out-of-plane magnetic anisotropies from the kagome plane should be determined by $K$ and $D$, respectively. Figures 1d and 1e schematically show spin-wave modes for a chiral AF metal. Modes I is an in-plane ($xy$) optical mode. While out-of-plane precession ($z$) is in-phase and is governed by $D$, in-plane precession is out-of-phase and is determined by $J$. Therefore, a typical energy scale ($\hbar\omega_I$) of the mode I can be estimated as $\sim\sqrt{JD}$. Mode II is collective precession-like motion, that is, in-plane acoustic mode. Similarly, a typical energy scale ($\hbar\omega_{II}$) of the mode II can be $\sim$ $\sqrt{KJ}$. As we discuss in detail below, while the resonant frequencies of modes I and II are different, typical time scales for the magnetization switching are identical, $(\Delta\omega_\mathrm{I})^{-1} \approx (\Delta\omega_\mathrm{II})^{-1} \sim (\alpha J)^{-1}\hbar$ and thus is much shorter by the factor of $K/J$ than the ferromagnetic case. Such an *exchange-enhanced* ultrafast damped precession should be available in both antiferromagnetic and ferrimagnetic metals and would be the most significant in a fully compensated case. However, such compensation of magnetic moments has been considered to make it impossible to observe the spin-dynamics in AF metals. This is because a signal amplitude is usually proportional to the net spin-magnetization in a detection method for the spin-dynamics e.g. the Faraday effect and the MOKE. Here we demonstrate the time-resolved ultrafast spin precession in the AF Weyl semimetal $Mn_3Sn$ by using MOKE induced by the magnetic octupole order. The spin precessions from the non-collinear texture induce the oscillations of MOKE and the Berry curvature in the momentum space.

In this study, bulk single-crystal $D0_{19}$-$Mn_3Sn$ has been employed (Supporting Information 2.). $Mn_3Sn$ has the hexagonal $Ni_3Sn$-type crystal structure consisting of an *ABAB* stacking of the kagome lattice of Mn atoms along with the [0001] axis. Red (blue) circles in Fig. 1a indicate Mn atoms in the *A*- (*B*-) plane of the kagome lattice. Below the Néel temperature of 430 K, an inverse triangular spin structure is stabilized by exchange and Dzyaloshinskii-Moriya interactions [10,39,40]. The inverse triangular spin structure possesses a uniform negative vector chirality of the in-plane Mn moments and is made of a ferroic ordering of cluster magnetic octupoles. The magnetic moments cant slightly in the (0001)-plane and produce a small net spontaneous magnetization [41], and can be reversed by an external magnetic field. Note that the appearance of the anomalous Hall effect [10], anomalous Nernst effect [13,14], and MOKE [31] is not induced by the spin-magnetization due to the canting but by the cluster magnetic octupole. As shown in Fig. 2a, the static MOKE signal has been confirmed in our crystal. Significantly, the polar MOKE signal, where the magnetic field is applied parallel to the kagome plane ($\boldsymbol{B}$ // $[2\bar{1}\bar{1}0]$), exhibits a clear hysteresis with a large Kerr rotation angle (~60 mdeg). This is three times larger than the previous report [31], and the difference may come from the optical interference effect [42]. In addition to the polar MOKE, the longitudinal signals, where the magnetic field is applied in $[01\bar{1}0]$ and [0001] directions, were also characterized (Fig. 2b). While a clear hysteresis curve was observed in $[01\bar{1}0]$, no signal was confirmed in [0001]. The observed magnetic anisotropy is consistent with the previous works, where the octupole polarization in $D0_{19}$-Mn$_3$Sn was detected via the measurements of the anomalous Hall effect [10], anomalous Nernst effect [13], and MOKE [31]. The Right (left) vertical axis in Fig. 2 shows the polar Kerr rotation angle measured with 660-nm continuous-wave (800-nm pulse) laser system. The difference in the Kerr rotation angle obtained by the 660-nm and the 800-nm laser systems is consistent with the MOKE spectroscopy reported previously [31]. Insets show magnetization hysteresis curves measured with the same magnetic field configuration as the MOKE measurements. From Fig. 2a inset, the spontaneous magnetization of Mn$_3$Sn is $9\times10^{-3}\mu_B$ per f.u.. If we assume a conventional ferromagnet (e.g. Fe, Co, and Ni), a possible Kerr rotation angle at zero magnetic field from the spontaneous magnetization would be as small as 0.2 mdeg with positive polarity [31]. Therefore, a large negative Kerr rotation angle at zero magnetic field can be hardly explained by the spin-magnetization due to canting.

The octupole has the same irreducible representation of $T_{1g}$ as the spin-magnetization, and can induce the MOKE. The results of multipole expansion of the antiferromagnetic structure of Mn$_3$Sn show that the octupole contributes to the expansion by more than 99.9% [32]. The previous first-principles calculation finds that the Kerr rotation angle has nearly no contribution from the net magnetization [31]. Moreover, a method to obtain a low-energy effective model of Mn$_3$Sn has been recently developed based on the cluster multipole theory [38]. The effective model presented in Ref. 38 well reproduces the results for the domain wall dynamics and for the coherent steady precession of spins obtained by using the original spin Hamiltonian (Eq. 1).

Our measurement of the time-resolved magneto-optical Kerr effect (TR-MOKE) was made by using an all-optical pump-probe method at room temperature (Fig. 3a and Supporting Information 3.). This method has been conventionally employed to detect spin precession for ferromagnetic metals [43] via Kerr effect and AF insulators [35-37] via the Faraday effect. For an AF metal, however, no report on time-resolved spin precession has been made to date. Pump and probe lights were configured almost perpendicular to the Mn$_3$Sn $(2\bar{1}\bar{1}0)$-surface. In addition, an external magnetic field normal to the surface (// $[2\bar{1}\bar{1}0]$) was applied during the measurements to direct the octupole polarization.

Figure 3b shows typical TR-MOKE results, where a magnetic field of 2 T was applied normal to the surface. First, a pump light induces a rapid increase in the Ker rotation angle of the probe light. Because Mn$_3$Sn is metallic, a pump pulse rapidly increases the electron temperature of the system. This rapid increase of electron temperature causes a significant decrease in the size of the order parameter, namely, the cluster magnetic octupole, similar to the case in ultrafast demagnetization in ferromagnetic metals [43,44]. After the excitation, a coherent spin precession starts with the aid of an effective magnetic field. As the cluster magnetic octupole is the order parameter that induces the MOKE signal as we discussed above, the spin precession is observed as the fluctuation of the cluster magnetic octupole. In Fig. 3B, the MOKE intensity starts to recover with delay time more than 0.3 ps, exhibiting a small but clear oscillation during the recovery. To further characterize the oscillating component, a non-oscillating component was estimated as a background (black solid curve) and subtracted from the raw data (Supporting Information 4.). The lower panel of Fig. 3b shows the TR-MOKE signals after subtracting the background. Figure 3c also shows typical TR-MOKE results (upper panel) and the analysis for the oscillating component (lower panel) in a relatively long-time range. The orange (blue) thick curve in the lower panel of Fig. 3b (3c) corresponds to the fitting to the equation, $\cos(\omega_{I(II)}t + \phi_0) \exp(-\alpha_{I(II)}\omega_{I(II)}t)$, which yields both the resonant frequency and effective damping constant at $B = 2$ T, i.e. $\omega_{I(II)}/2\pi = 0.86$ THz (18 GHz) and $\alpha_{I(II)} = 0.02$ (1.0), respectively. The orange (blue) thin curves represent the envelope functions of the fit, $\pm\exp(-\alpha_{I(II)}\omega_{I(II)}t)$. Here, $t$ is the delay time. The oscillation frequencies ($\omega_I$, $\omega_{II}$) as a function of the external magnetic field are displayed in Fig. 3d, which will be discussed later in detail. The oscillating behaviour is most significant when the magnetic field is normal to the $Mn_3Sn$ surface. Because the TR-MOKE signal is proportional to the surface normal polarization of the cluster magnetic octupoles, the oscillating Kerr signal should not come from a precession-like motion but from a change of size in the octupole order parameter. The change of size in the octupole order parameter originates from the spin precession of each magnetic moment. Specifically for the mode I, the signal-to-noise ratio of the TR-MOKE significantly reduces with decreasing the magnetic field strength below 2 T. This is because an external magnetic field is necessary to create an effective magnetic field to drive a coherent precession (Supporting Information 5.).

Two oscillation modes I and II are found through the analyses shown in Figs. 3b and 3c, which should come from the oscillation of the cluster magnetic octupoles (Supporting Information 4.). To analyze the modes, we have derived the analytical solution to estimate the resonant frequencies from Eq. 1. Note that Eq. 1 includes the in-plane exchange interaction $J$ in the kagome lattice, but not the inter-plane exchange interaction, and thus is only valid for the modes where precessional motions in different kagome-planes in-phase. The equation of motion for low energy magnetic excitation can be derived from Eq. 1 as the following sine-Gordon equation (Supporting Information 6.):

$$\frac{\hbar}{2\sqrt{3}\left(D+\sqrt{3}J\right)S}\ddot{\phi}+\alpha\dot{\phi}-\frac{\left(\sqrt{3}D+J\right)S}{2\hbar}a_{\text{lat}}^2\partial^2\phi+\frac{KS}{2\hbar}\sin 2\phi=0, \quad (2)$$

where $\hbar$, $\alpha$, and $a_{\text{lat}}$ indicate the reduced Planck constant, damping constant, and the lattice constant of the nearest neighbor Mn atoms, respectively. Here $S$ is the size of spin angular momentum ~1.5 for the Mn magnetic moment ~$3\mu_B$, $\phi$ refers to in-plane precession angle of Mn magnetic moment, and $\alpha$ is identical to the Gilbert damping constant in the Landau-Lifshitz-Gilbert equation. From Eq. 2, resonant frequencies for modes I and II can be estimated as follows:

$$\hbar\omega_{\text{I}} = S\sqrt{6\sqrt{3}\left(\sqrt{3}D+J\right)D}, \quad (3)$$

$$\hbar\omega_{\text{II}} = S\sqrt{2\sqrt{3}K\left(D+\sqrt{3}J\right)}. \quad (4)$$

Here the resonant frequencies for the modes I and II ($\omega_{\text{I}}$ and $\omega_{\text{II}}$) correspond to the optical and collective precession-like modes depicted in Figs. 1d and 1e, respectively. The damped oscillation is often expressed by $\exp(-\alpha_{I(II)}\omega_{I(II)}t)$ by introducing a phenomenological effective damping constant ($\alpha_I$, $\alpha_{II}$), which is expressed as

$$\hbar\omega_I\alpha_I = \hbar\omega_{II}\alpha_{II} = 2\sqrt{3}S\left(D+\sqrt{3}J\right)\alpha. \qquad (5)$$

Note that the effective damping ($\alpha_I$, $\alpha_{II}$) is not identical to the Gilbert damping constant ($\alpha$) defined in Eq. 2. The Gilbert damping constant should be determined from the damping rate (Eq. 5), which is independent of the resonant frequency ($\omega_I$, $\omega_{II}$). While these two damping constants are almost identical for the case of a ferromagnet, a large difference between $\omega_I$ and $\omega_{II}$ induces large deviation of the effective damping constant from the Gilbert damping constant in $Mn_3Sn$. This is a unique property of the octupole oscillation dynamics in a chiral AF metal and should be distinguished from spin-magnetization dynamics in a ferromagnet. Very recently, related discussion for the damping has been made in the magnetic domain-wall dynamics in the ferrimagnetic GeFeCo [45]. Here, the field dependence of the resonant frequencies (Fig. 3d) is understood as follows. The resonant frequency of the optical mode ($\omega_I$) is determined by interactions between Mn atoms, and thus is insensitive to an external magnetic field. However, the resonant frequency of the collective precession-like mode ($\omega_{II}$) is not. This is because the cluster magnetic octupole can couple with an external magnetic field via spontaneous magnetization due to canting. From the field dependence of the resonant frequencies (Fig. 3d), $\omega_{II}/2\pi$ at $B = 0$ is determined to be 13.7±1.5 GHz.

To characterize the dynamics of the cluster magnetic octupole, the six-fold magnetic anisotropy $K$ was determined from the torque measurements (Supporting Information 7.). Figure 4a shows the result of out-of-plane rotation (*y*-axis) in terms of the plane consisting of the kagome lattice in $Mn_3Sn$ (*xy*-plane) by rotating the direction of the magnetic field (**B**) from $[2\bar{1}\bar{1}0]$ ($\phi_{B1} = 0°$) to $[0001]$ ($\phi_{B1} = 90°$). The solid and open circles indicate different rotation directions. The results allow the estimation of the saturation magnetization ($M_S$), as displayed in the inset of Fig. 4a. The saturation magnetization should correspond to the spontaneous magnetization due to spin canting at zero field, determined by $D$ and $J$, and should be distinguished from the magnetization components due to the spin canting induced by the application of an external magnetic field. Here the saturation magnetization is estimated to be $11.3 \times 10^{-3} \mu_B$ ($10.0 \times 10^{-3} \mu_B$) per formula unit (f.u.) at $B = 0$ T (9 T), and is comparable to the magnetization at $B = 0$ ($9 \times 10^{-3} \mu_B$ per f.u. from Fig. 2a inset). On the other hand, the in-plane rotation (z-axis) measurements of the torque was made by rotating the direction of the magnetic field (**B**) from $[2\bar{1}\bar{1}0]$ ($\phi_{B2} = 0°$) to $[01\bar{1}0]$ ($\phi_{B2} = 90°$) within the kagome-lattice plane (Fig. 4b). The analysis of the result estimates an energy barrier height from the six-fold in-plane magnetic anisotropy energy ($K_6/18$, Supporting Information 7.) to be $= 3.1 \times 10^2$ Jm$^{-3}$, consistent with the value reported in the previous work ($2.2 \times 10^2$ Jm$^{-3}$) [46]. Here, the obtained in-plane magnetic anisotropy energy corresponds to $K = 3.1 \times 10^{-4}$ meV from Eq. 1.

In principle, six spin-wave modes should be confirmed in Mn$_3$Sn as the unit cell consists of six Mn atoms. Previous neutron scattering studies [47, 48] have revealed the three modes in the low energy region ($\hbar\omega < 20$ meV) and the other three modes in the high energy region ($\hbar\omega \approx 100$ meV) at **q** = 0, where **q** is the momentum vector. The low and high energy modes are the ones where precessional motions in different kagome-planes are in-phase and out-of-phase, respectively. Here, both the modes I and II in this study correspond to the low energy modes. From TR-MOKE results (Fig. 3B), the energy of the mode I is 3.6 meV (0.86 THz). However, this is only 26% of the value obtained in the neutron scattering (~14 meV) [47]. This strongly suggests that the size of the magnetic moment (see $S$ in Eq. 3) is reduced. The Néel temperature of $Mn_3Sn$ is not so far from room temperature [39] that the typical temperature during the TR-MOKE measurements might be close to the Néel temperature due to heating effect, which may reduce the size of the magnetic moment. In fact, it has been reported that the anomalous Hall effect and MOKE are significantly suppressed by heating $Mn_3Sn$ [49]. When $S \sim 0.4$ (= 1.5×26%) is employed in Eqs. 3 and 4, $J$ and $D$ are estimated to be ~10 meV and ~0.7 meV, respectively, which is in good agreement with the previous studies [47, 48, 50]. Gilbert damping constant of $Mn_3Sn$, which is estimated for the first time, is found to be $\alpha = 0.0005$ from Eq. 5. The Gilbert damping constant of $Mn_3Sn$ is as small as the theoretical prediction for Mn-Ge [51] and Mn-Ga [52] alloys (0.0005-0.001).

As discussed, the dynamics characteristic of the cluster magnetic octupoles is an ultrafast damped oscillation due to the exchange-interaction (Eq. 5). Interestingly, the exchange-interaction dramatically increases the effective damping ($\alpha_{II} = 1.0$). Thus, the typical switching time can be ultrafast ($1/\omega_I\alpha_I \approx 1/\omega_{II}\alpha_{II} \sim 9$ ps) although the resonant frequency for the collective precession-like motion is relatively slow ($\omega_{II}/2\pi = 13$ GHz). As discussed, $S \sim 0.4$ was employed for the aforementioned analysis. If the sample temperature could be held much lower than the Néel temperature ($S = 1.5$), the switching time would be reduced to ~2 ps. From the sine-Gordon equation (Eq. 2), the Néel type-domain wall velocity in $Mn_3Sn$ can be estimated as fast as 3 km/s (12 km/s) for $S = 0.4$ (1.5) (Supporting Information 8.), which is greater than the recently reported value for the ferrimagnetic systems in the vicinity of the angular momentum compensation temperature [53-55].

For device applications, not only dynamic but also static properties of the cluster magnetic octupoles would be important. As discussed, energy barrier height ($K_6/18$) for the six-fold magnetic anisotropy energy of Mn$_3$Sn is $3.1\times10^2$ J m$^{-3}$, which corresponds to the in-plane magnetic anisotropy field of 3.8 T (= $K_6/M_S$). If we employ 10-nm-thick Mn$_3$Sn, the areal magnetic anisotropy energy should correspond to 3 µJ/m$^2$. To obtain a thermal stability factor of 60 at 300 K, which is defined as magnetic anisotropy energy divided by thermal fluctuation energy ($k_B T$), relatively large magnetic cell as 320 nm in diameter ($8\times10^5$ nm$^3$) is needed. However, such small magnetic anisotropy energy ensures that the octupole polarization can be efficiently controlled by electric current- or voltage-driven torque [56].

To conclude, our work demonstrates the giant effective damping of the octupole dynamics in an antiferromagnetic Weyl semimetal. The exchange interaction significantly increases the damping rate for the collective precession-like mode, which is directly related to a typical switching time in device operation [57]. The introduction of the key concept of *cluster magnetic octupole*, instead of *spin-magnetization* in ferromagnets, has provided the basis for carrying out the experiments by using the conventional schemes developed for ferromagnetic spintronics. Thus, our observation will certainly foster the development of the topological spintronics.

FIGURES:

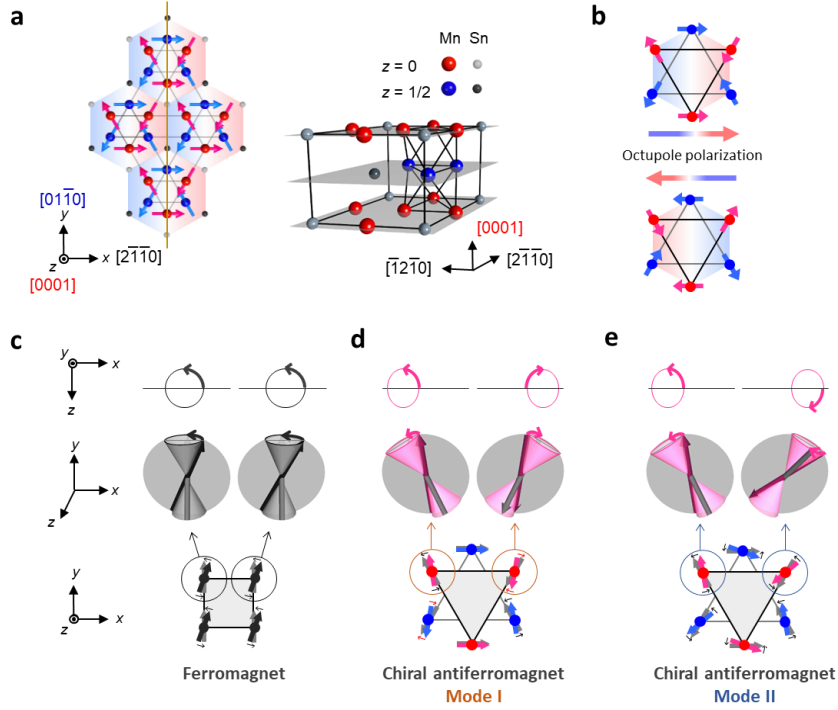

**Figure 1.** Cluster magnetic octupole and its oscillation modes in $D0_{19}$-Mn$_3$Sn. (a) Spin and crystal structures of Mn$_3$Sn. Different colors are used to denote Mn and Sn atoms in the $z = 0$ and $z = 1/2$ planes. The inverse triangular spin-structure can be viewed as a ferroic ordering of cluster magnetic octupole, which possesses the same symmetry as the spin-magnetization (e.g. as indicated by a mirror plane (orange line)). (b) Magnetic cluster octupole units with different octupole polarization directions. (c) Spin-dynamics in a ferromagnet with uniaxial magnetic anisotropy ($K$). Resonant frequency ($\hbar\omega$) and typical switching time ($\Delta\omega$)$^{-1}$ can be ~$K$ and ~$(2\alpha K)^{-1}\hbar$, respectively, where $\alpha$ is the Gilbert damping constant. (d), (e) Spin-dynamics in a chiral antiferromagnet. Spin-wave modes I and II correspond to optical and collective precession-like motions and possess resonant frequencies of $\omega_{\mathrm{I}} \approx \sqrt{JD}\hbar^{-1}$ and $\omega_{\mathrm{II}} \approx \sqrt{KJ}\hbar^{-1}$, respectively. $D$ and $J$ are Dzyaloshinskii-Moriya interaction and exchange interaction energy scales, respectively. Typical switching time for both modes can be estimated as $(\Delta\omega_\text{I})^{-1} \approx (\Delta\omega_\text{II})^{-1} \sim (\alpha J)^{-1}\hbar$.

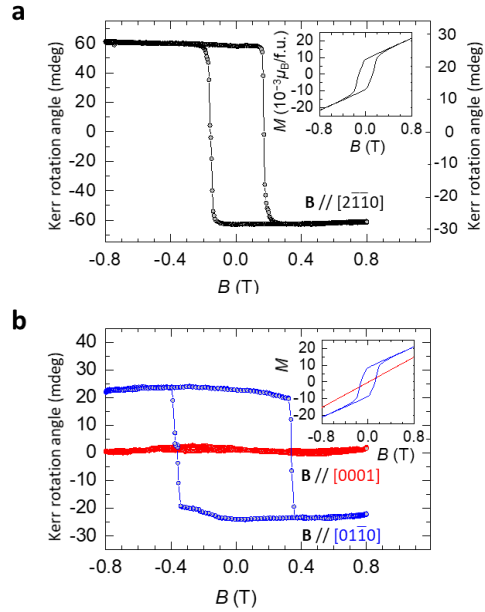

**Figure 2.** Static hysteresis loops of the MOKE. (a) Magnetic field dependence of the polar MOKE result. (b) Magnetic field dependence of the longitudinal MOKE results. Left (right) vertical axis shows Kerr rotation angle measured with 660-nm continuous-wave (800-nm pulse) laser system. Insets show the corresponding magnetization hysteresis ($M$ in $10^{-3}\mu_B$ per f.u.) obtained as a function of the magnetic field with the same configuration as the MOKE in the main panel. MOKE signal does not come from the spin-magnetization due to canting but from cluster magnetic octupole as discussed in the main text and Ref. 31.

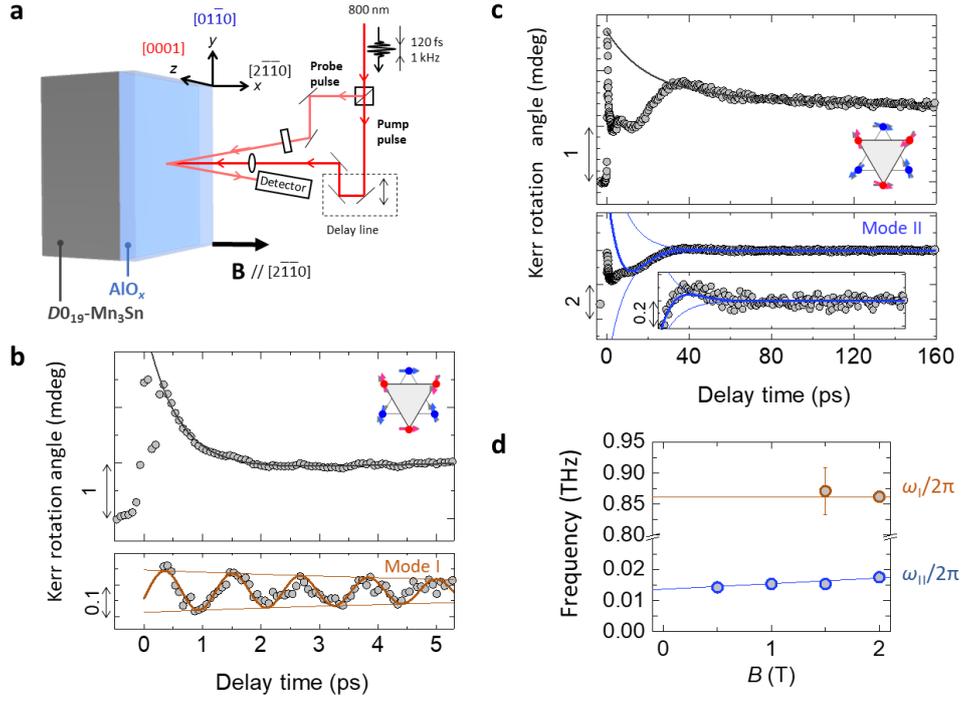

**Figure 3.** Time-resolved oscillation of cluster magnetic octupole. (a) Schematic diagram of the time-resolved polar MOKE measurements (TR-MOKE). (b), (c) TR-MOKE signals of Mn$_3$Sn under a magnetic field of 2 T in the $[2\bar{1}\bar{1}0]$ direction. Backgrounds, indicated by black curves in upper panels, were subtracted from the raw data and plotted in the lower panels. Orange (blue) thick curve corresponds to the fit, which yields the resonant frequency and effective damping constant, $\omega_{I(II)}/2\pi$ = 0.86 THz (18 GHz) and $\alpha_{I(II)}$ = 0.02 (1.0), respectively. (d) Magnetic field dependence of the oscillation frequency for the modes I (orange open circle) and II (blue open circle).

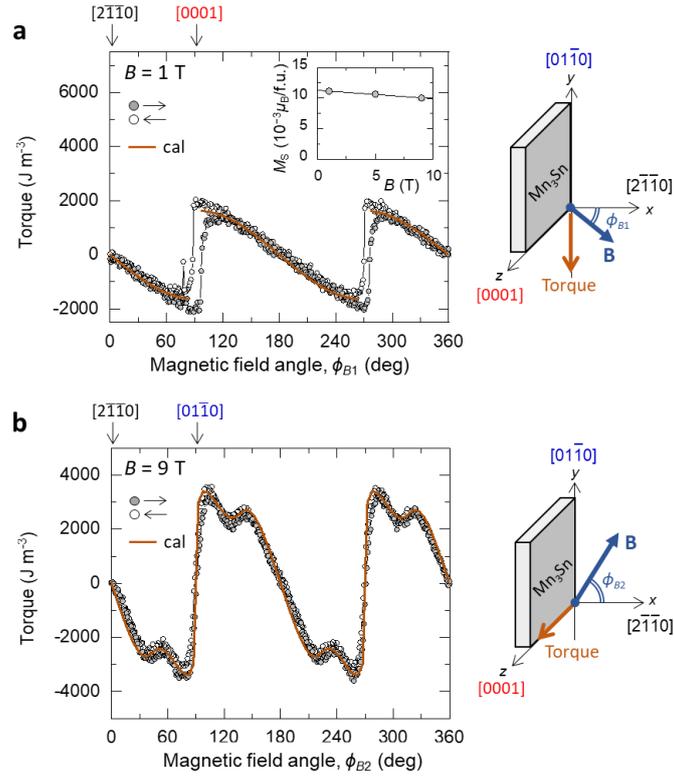

**Figure 4.** Magnetic anisotropy. (a) Results of the torque measurements for the out-of-plane magnetic field rotation, namely, upon changing the direction of the magnetic field $B$ from $[2\bar{1}\bar{1}0]$ ($\phi_{B1}$ = 0°) to [0001] ($\phi_{B1}$ = 90°). The inset shows the saturation magnetization ($M_S$) as a function of an external magnetic field $B$. (b) Results of the torque measurements for the in-plane magnetic field rotation, namely, upon changing the direction of the magnetic field $B$ from $[2\bar{1}\bar{1}0]$ ($\phi_{B2}$ = 0°) to [0001] ($\phi_{B2}$ = 90°). Solid and open circles represent the torque obtained with increasing and decreasing rotation angles of $\phi_{B1}$ and $\phi_{B2}$.

AUTHOR INFORMATION

**Corresponding Author**

miwa@issp.u-tokyo.ac.jp

nomoto@ap.t.u-tokyo.ac.jp.

**Author Contributions**

S. Miwa, S. I., S. S. and S. Mizukami conducted the TR-MOKE measurements. T. N. and R. A. conducted the theoretical study. T. T. and Y. O. conducted the torque measurements. S. Miwa, T. H., M. I., and S. S. prepared the sample. T. H., S. Miwa and S. S. conducted the magnetization measurements. S. Miwa and S. N. conceived and designed the experiments, and wrote the paper. All authors discussed the results and commented on the manuscript.

**Funding Sources**

This work was partially supported by JSPS-KAKENHI (Nos. JP18H03880, JP26103002, JP19H05825, JP19H00650, JP16H06345, JP15H05882, and JP15H05883), JST-CREST (No. JPMJCR18T3), JST-Mirai Program(JPMJMI20A1) and the Spintronics Research Network of Japan (Spin-RNJ).

**Notes**

ACKNOWLEDGEMENT

We thank K. Kondou of RIKEN, T. Taniguchi, T. Nakano, Y. Hibino, T. Nozaki, M. Konoto, K. Yakushiji, and A. Fukushima of AIST, R. Matsunaga of the University of Tokyo for discussion, Y. Suzuki, and M. Goto of Osaka University, Y. Kato of the

REFERENCES

1. Nielsen, H. B.; Ninomiya M. The Adler-Bell Jackiw anomaly and Weyl fermions in a crystal. *Phys. Lett. B* **1983**, *130*, 389.

2. Wan X.; Turner, A. M.; Vishwanath, A.; Savrasov, S. Y. Topological semimetal and Fermi-arc surface states in the electronic structure of pyrochlore iridates. *Phys. Rev. B* **2011**, *83*, 205101.

3. Burkov, A. A.; Balents, L. Weyl semimetal in a topological insulator multilayer. *Phys. Rev. Lett.* **2011**, *107*, 127205.

4. Armitage, N. P.; Mele, E. J.; Vishwanath, A. Weyl and Dirac semimetals in three-dimensional solids. *Rev. Mod. Phys.* **2018**, *90*, 015001.

5. Yang, K. Y.; Lu, Y. M.; Ran, Y. Quantum Hall effects in a Weyl semimetal: possible application in pyrochlore iridates. *Phys. Rev. B* **2011**, *84*, 075129.

6. Zyuzin, A. A.; Burkov, A. A.; Topological response in Weyl semimetals and the chiral anomaly. *Phys. Rev. B* **2012**, *86*, 115133.

7. Son, D. T.; Spivak, B. Z.; Chiral anomaly and classical negative magnetoresistance of Weyl metals. *Phys. Rev. B* **2013**, *88*, 104412.

8. Zhong, S.; Orenstein, J.; Moore, J. E.; Optical gyrotropy from axion electrodynamics in momentum space. *Phys. Rev. Lett.* **2015**, *115*, 117403.

9. Xiong, J.; Kushwaha, S. K.; Liang, T.; Krizan, J. W.; Hirschberger, M.; Wang, W.; Cava, R. J.; Ong, N. P. Evidence for the chiral anomaly in the Dirac semimetal Na$_3$Bi. *Science* **2015**, *350*, 413.

10. Nakatsuji, S.; Kiyohara, N.; Higo, T. Large anomalous Hall effect in a

non-collinear antiferromaget at room temperature. *Nature* **2015**, *527*, 212.

11. Sharma, G.; Goswami, P.; Tewari, S. Nernst and magnetothermal conductivity in a lattice model of Weyl fermions. *Phys. Rev. B* **2016**, *93*, 035116.

12. Hirschberger, M.; Kushwaha, S.; Wang, Z.; Gibson, Q.; Liang, S.; Belvin, C. A.; Bernevig, B. A.; Cava, R. J.; Ong, N. P. The chiral anomaly and thermopower of Weyl fermions in the half-Heusler GdPtBi. *Nat. Mater.* **2016**, *15*, 1161.

13. Ikhlas, M.; Tomita, T.; Koretsune, T.; Suzuki, M.-T.; Nishio-Hamane, D.; Arita, R.; Otani, Y.; Nakatsuji, S. Large anomalous Nernst effect at room temperature in a chiral antiferromagnet. *Nat. Phys.* **2017**, *13*, 1085.

14. Li, X.; Xu, L.; Ding, L.; Wang, J.; Shen, M.; Lu, X.; Zhu, Z.; Behnia, K. Anomalous Nernst and Righi–Leduc effects in $Mn_3Sn$: Berry curvature and entropy flow. *Phys. Rev. Lett.* **2017**, *119*, 056601.

15. Kuroda, K.; Tomita, T.; Suzuki, M.-T.; Bareille, C.; Nugroho, A. A.; Goswami, P.; Ochi, M.; Ikhlas, M.; Nakayama, M.; Akebi, S.; Noguchi, R.; Ishii, R.; Inami, N.; Ono, K.; Kumigashira, H.; Varykhalov, A.; Muro, T.; Koretsune, T.; Arita, R.; Shin, S.; Kondo, T.; Nakatsuji, S. Evidence for magnetic Weyl fermions in a correlated metal. *Nat. Mater.* **2017**, *16*, 1090.

16. Sakai, A.; Mizuta, Y. P.; Nugroho, A. A.; Sihombing, R.; Koretsune, T.; Suzuki, M. T.; Takemori, N.; Ishii, R.; Nishio-Hamane, D.; Arita, R.; Goswami, P.; Nakatsuji, S. Giant anomalous Nernst effect and quantum-critical scaling in a ferromagnetic semimetal. *Nat. Phys*. **2018**, *14*, 1119.

17. Liu, E.; Sun, Y.; Kumar, N.; Muechler, L.; Sun, A.; Jiao, L.; Yang, S. Y.; Liu, D.; Liang, A.; Xu, Q.; Kroder, J.; Süß, V.; Borrmann, H.; Shekhar, C.; Wang, Z.; Xi, C.; Wang, W.; Schnelle, W.; Wirth, S.; Chen, Y.; Goennenwein, S. T. B.; Felser, C.

Giant anomalous Hall effect in a ferromagnetic kagome-lattice semimetal. *Nat. Phys*. **2018**, *14*, 1125.

18. Wang, Q.; Xu, Y.; Lou, R.; Liu, Z.; Li, M.; Huang, Y.; Shen, D.; Weng, H.; Wang, S.; Lei, H. Large intrinsic anomalous Hall effect in half-metallic ferromagnet $Co_3Sn_2S_2$ with magnetic Weyl fermions. *Nat. Commun*. **2018**, *9*, 3681.

19. Jungwirth, T.; Marti, X.; Wunderlich, J. Antiferromagnetic spintronics. *Nat. Nanotechnol*. **2016**, *11*, 231.

20. Baltz, V.; Manchon, A.; Tsoi, M.; Moriyama, T.; Ono, T.; Tserkovnyak, Y. Antiferromagnetic spintronics. *Rev. Mod. Phys*. **2018**, *90*, 015005.

21. Němec, P.; Fiebig, M.; Kampfrath, T.; Kimel, A. V. Antiferromagnetic opto-spintronics. *Nat. Phys*. **2018**, *14*, 229.

22. Marti, X.; Fina, I.; Frontera, C.; Liu, Jian; Wadley, P.; He, Q.; Paull, R. J.; Clarkson, J. D.; Kudrnovský, J.; Turek, I.; Kuneš, J.; Yi, D.; Chu, J-H.; Nelson, C. T.; You, L.; Arenholz, E.; Salahuddin, S.; Fontcuberta, J.; Jungwirth, T.; Ramesh, R. Room-temperature antiferromagnetic memory resistor. *Nat. Mater*. **2014**, *13*, 367.

23. Wadley, P.; Howells, B.; J. Železný, Andrews, C.; Hills, V.; Campion, R. P.; Novák, V.; Olejník, K.; Maccherozzi, F.; Dhesi, S. S.; Martin, S. Y.; Wagner, T.; Wunderlich, J.; Freimuth, F.; Mokrousov, Y.; Kuneš, J.; Chauhan, J. S.; Grzybowski, M. J.; Rushforth, A. W.; Edmonds, K. W.; Gallagher, B. L.; Jungwirth, T. Electrical switching of an antiferromagnet. *Science* **2016**, *351*, 587.

24. Wadley, P., Reimers, S., Grzybowski, M. J.; Andrews, C.; Wang, M.; Chauhan, J. S. ; Gallagher, B. L.; Campion, R. P.; Edmonds, K. W.; Dhesi, S. S.; Maccherozzi, F.; Novak, V.; Wunderlich, J.; Jungwirth, T. Current polarity-dependent manipulation of antiferromagnetic domains. *Nat. Nanotechnol.* **2018**, *13*, 362.

25. Moriyama, T.; Oda, K.; Ohkochi, T.; Kimata, M.; Ono, T. Spin torque control of antiferromagnetic moments in NiO. *Sci. Rep.* **2018**, *8*, 14167.

26. Baldrati, L.; Gomonay, O.; Ross, A.; Filianina, M.; Lebrun, R.; Ramos, R.; Leveille, C.; Fuhrmann, F.; Forrest, T. R.; Maccherozzi, F.: Valencia, S.; Kronast, F.; Saitoh, E.; Sinova, J.; Kläui, M. Mechanism of Néel order switching in antiferromagnetic thin films revealed by magnetotransport and direct imaging. *Phys. Rev. Lett.* **2019**, *123*, 177201.

27. Reichlova, H.; Janda, T.; Godinho, J.; Markou, A.; Kriegner, D.; Schlitz, R.; Zelezny, J.; Soban, Z.; Bejarano, M.; Schultheiss, H.; Nemec, P.; Jungwirth, T.; Felser, C.; Wunderlich, J.; Goennenwein, S. T. B. Imaging and writing magnetic domains in the non-collinear antiferromagnet $Mn_3Sn$. *Nat. Commun.* **2019**, *10*, 5459.

28. Cheong, S.-W.; Fiebig, M.; Wu, W.; Chapon, L.; Kiryukhin, V. Seeing is believing: visualization of antiferromagnetic domains. *npj Quant. Mater.* **2020**, *5*, 3

29. Saidl, V.; Němec, P.; Wadley, P.; Hills, V.; Campion, R. P.; Novák, V.; Edmonds, K. W.; Maccherozzi, F.; Dhesi, S. S.; Gallagher, B. L.; Trojánek, F.; Kuneš, J.; Železný, J.; Malý, P.; Jungwirth, T. Optical determination of the Néel vector in a CuMnAs thin-film antiferromagnet. *Nat. Photon*. **2017**, *11*, 91.

30. Thielemann-Kühn, N.; Schick, D.; Pontius, N.; Trabant, C.; Mitzner, R.; Holldack, K.; Zabel, H.; Föhlisch, A.; Schüßler-Langeheine, C. Ultrafast and energy-efficient quenching of spin order: antiferromagnetism beats ferromagnetism. *Phys. Rev. Lett*. **2017**, *119*, 197202.

31. Higo, T.; Man, H.; Gopman, D. B.; Wu, L.; Koretsune, T.; van 't Erve, O. M. J.; Kabanov, Y. P.; Rees, D.; Li, Y.; Suzuki, M.-T.; Patankar, S.; Ikhlas, M.; Chien, C. L.; Arita, R.; Shull, R. D.; Orenstein, J.; Nakatsuji, S. Large mageto-opitical Kerr effect and imaging of magnetic octupole domains in an antiferromagnetic metal. *Nat. Photon*. **2018**, *12*, 73.

32. Suzuki, M.-T.; Koretsune, T.; Ochi, M.; Arita, R. Cluster multipole theory for anomalous Hall effect in antiferromagnets. *Phys. Rev. B* **2017**, *95*, 094406.

33. Olejník, K.; Seifert, T.; Kašpar, Z.; Novák, V.; Wadley, P.; Campion, R. P.; Baumgartner, M.; Gambardella, P.; Němec, P.; Wunderlich, J.; Sinova, J.; Kužel, P.; Müller, M.; Kampfrath, T.; Jungwirth, T. Terahertz electrical writing speed in an antiferromagnetic memory. *Sci. Adv*. **2018**, *4*, eaar3566.

34. Ju, G.; Hohlfeld, J.; Bergman, B.; van de Veerdonk, R. J. M.; Mryasov, O. N.; Kim, J.-Y.; Wu, X.; Weller, D.; Koopmans, B. Ultrafast generation of ferromagnetic order via a laser-induced phase transformation in FeRh thin film. *Phys. Rev. Lett*. **2004**, *93*, 197403.

35. Kimel, A. V.; Kirilyuk, A.; Usachev, P. A.; Pisarev, R. V.; Balbashov, A. M.; Rasing, T. Ultrafast non-thermal control of magnetization by instantaneous photomagnetic pulses. *Nature* **2005**, *435*, 655.

36. Satoh, T.; Cho, S.-J.; Iida, R.; Shimura, T.; Kuroda, K.; Ueda, H.; Ueda, Y.; Ivanov, B. A.; Nori, F.; Fiebig, M. Spin oscillation in antiferromagnetic NiO triggered by circularly polarized light. *Phys. Rev. Lett*. **2010**, *105*, 077402.

37. Kampfrath, T.; Sell, A.; Klatt, G.; Pashkin, A.; Mährlein, S.; Dekorsy, T.; Wolf, M.; Fiebig, M.; Leitenstorfer, A.; Huber, R. Coherent terahertz control of antiferromagnetic spin waves. *Nat. Photon*. **2011**, *5*, 31.

38. Nomoto, T.; Arita, R. Cluster multipole dynamics in noncollinear antiferromagnets. *Phys. Rev. Research* **2020**, *2*, 012045(R).

39. Krén, E.; Paitz, J.; Zimmer, G.; Zsoldos, É. Study of the magnetic phase transition in the $Mn_3Sn$ phase. *Physica B* **1975**, *80*, 226.

40. Tomiyoshi, S.; Yamaguchi, Y. Magnetic structure and weak ferromagntism of $Mn_3Sn$ studied by polarized neutron diffraction. *J. Phys. Soc. Jpn*. **1982**, *51*, 2478.

41. Nagamiya, T.; Tomiyoshi, S.; Yamaguchi, Y. Triangular spin configuration and weak ferromagnetism of $Mn_3Sn$ and $Mn_3Ge$. *Solid State. Commun.* **1982**, *42*, 385.

42. Ahn, K. Y.; Fan, G. J. Kerr effect enhancement in ferromagnetic films. *IEEE Trans. Magn*. **1966**, *2*, 678.

43. Beauprepaire, E.; Merle, J.-C.; Daunois, A.; Bigot, J.-Y. Ultrafast spin dynamics in ferromagnetic nickel. *Phys. Rev. Lett.* **1996**, *76*, 4250.

44. Kirilyuk, A.; Kimel, A. V.; Rasing, T. Ultrafast optical manipulation of magnetic order. *Rev. Mod. Phys.* **2010**, *82*, 2731.

45. Kim, D.-H.; Okuno, T.; Kim, S. K.; Oh, S.-H.; Nishimura, T.; Hirata, Y.; Futakawa, Y.; Yoshikawa, H.; Tsukamoto, A.; Tserkovnyak, Y.; Shiota, Y.; Moriyama, T.; Kim, K.-J.; Lee, K.-J.; Ono, T. Low magnetic damping of ferrimagnetic GdFeCo alloys. *Phys. Rev. Lett*. **2019**, *122*, 127203.

46. Duan, T. F.; Ren, W. J.; Liu, W. L.; Li, S. J.; Liu, W.; Zhang, Z. D. Magnetic anisotropy of single-crystalline $Mn_3Sn$ in triangular and helix-phase states. *Appl. Phys. Lett.* **2015**, *107*, 082403.

47. Cable, J. W.; Wakabayashi, N.; Radhakrishna, P. Magnetic excitation in the triangular antiferromagnets $Mn_3Sn$ and $Mn_3Ge$. *Phys. Rev. B* **1993**, *48*, 6159.

48. Park, P.; Oh, J.; Uhlířová, K.; Jackson, J.; Deák, A.; Szunyogh, L.; Lee, K. H.;

Cho, H.; Kim, H.-L.; Walker, H. C.; Adroja, D.; Sechovský, V.; Park, J.-G. Magnetic excitations in non-collinear antiferromagnetic Weyl semimetal Mn$_3$Sn. *NPG Quant. Mater*. **2018**, *3*, 63.

49. Balk, A. L.; Sung, N. H.; Thomas, S. M.; Rosa, P. F. S.; McDonald, R. D.; Thompson, J. D.; Bauer, E. D.; Ronning, F.; Crooker, S. A. Comparing the anomalous Hall effect and the magneto-optical Kerr effect though angiferromagnetic phase transitions in Mn$_3$Sn. *Appl. Phys. Lett.* **2019**, *114*, 032401.

50. Liu, J.; Balents, L. Anomalous Hall effect and topological defects in antiferromagnetic Weyl semimetals: Mn$_3$Sn/Ge. *Phys. Rev. Lett.* **2017**, *119*, 087202.

51. Mizukami, S.; Sakuma, A.; Sugihara, A.; Kubota, T.; Kondo, Y.; Tsuchiya, H.; Miyazaki, T. Tetragonal D0$_{22}$ Mn$_{3+x}$Ge epitaxial film grown on MgO(100) with a large perpendicular magnetic anisotropy. *Appl. Phys. Express* **2013**, *6*, 123002.

52. Mizukami, S.; Wu, F.; Sakuma, A.; Walowski, J.; Watanabe, D.; Kubota, T.; Zhang, X.; Naganuma, H.; Oogane, M.; Ando, Y.; Miyazaki, T. Long-lived ultrafast spin precession in manganese alloys films with a large perpendicular magnetic anisotropy. *Phys. Rev. Lett.* **2011**, *106*, 117201.

53. Kim, K.-J.; Kim, S. K.; Hirata, Y.; Oh, S.-H.; Tono, T.; Kim, D.-H.; Okuno, T.; Ham, W. S.; Kim, S.; Go, G.; Tserkovnyak, Y.; Tsukamoto, A.; Moriyama, T.; Lee, K.-J.; Ono, T. Fast domain wall motion in the vicinity of the angular momentum compensation temperature of ferrimagnets. *Nat. Mater.* **2017**, *16*, 1187.

54. Caretta, L.; Mann, M.; Büttner, F.; Ueda, K.; Pfau, B.; Günther, C. M.; Hessing, P.; Churikova, A.; Klose, C.; Schneider, M.; Engel, D.; Marcus, C.; Bono, D.;

Bagschik, K.; Eisebitt, S.; Beach, G. S. D. Fast current-driven domain walls and small skyrmions in a compensated ferrimagnet. *Nat. Nanotechnol.* **2018**, *13*, 1154.

55. Cai, K.; Zhu, Z.; Lee, J. M.; Mishra, R.; Ren, L.; Pollard, S. D.; He, P.; Liang, G.; Teo, K. L.; Yang, H. Ultrafast and energy-efficient spin-orbit torque switching in compensated ferrimagnets. *Nat. Electron.* **2020**, *3*, 37.

56. Locatelli, N.; Cros, V.; Grollier, J. Spin-torque building blocks. *Nat. Mater.* **2014**, *13*, 11.

57. Tsai, H.; Higo, T.; Kondou, K.; Nomoto, T.; Sakai, A.; Kobayashi, A.; Nakano, T.; Yakushiji, K.; Arita, R.; Miwa, S.; Otani, Y.; Nakatsuji, S. Electrical manipulation of a topological antiferromagnetic state. *Nature* **2020**, *580*, 608.

Supporting Information

# Giant Effective Damping of Octupole Oscillation in an Antiferromagnetic Weyl Semimetal

Shinji Miwa[1,2,3*], Satoshi Iihama[4,5], Takuya Nomoto[3,6**], Takahiro Tomita[1,3], Tomoya Higo[1,3], Muhammad Ikhlas[1], Shoya Sakamoto[1], YoshiChika Otani[1,2,3,7], Shigemi Mizukami[4,5,8], Ryotaro Arita[3,6,7], and Satoru Nakatsuji[1,2,3,9]

[1]The Institute for Solid State Physics, The University of Tokyo, Kashiwa, Chiba 277-8581, Japan

[2]Trans-scale Quantum Science Institute, The University of Tokyo, Bunkyo, Tokyo 113-0033, Japan

[3]CREST, Japan Science and Technology Agency (JST), Kawaguchi, Saitama 332-0012, Japan

[4]Advanced Institute for Materials Research, Tohoku University, Sendai, Miyagi 980-8577, Japan

[5]Center for Spintronics Research Network (CSRN), Tohoku University, Sendai, Miyagi 980-8577, Japan

[6]Department of Applied Physics, The University of Tokyo, Tokyo 113-8656, Japan

[7]RIKEN, Center for Emergent Matter Science (CEMS), Saitama 351-0198, Japan

[8]Center for Science and Innovation in Spintronics (CSIS), Tohoku University, Sendai, Miyagi 980-8577, Japan

[9]Department of Physics, The University of Tokyo, Tokyo 113-0033, Japan

*miwa@issp.u-tokyo.ac.jp

**nomoto@ap.t.u-tokyo.ac.jp.

## 1. Resonant frequency and spectral linewidth

When we treat the spherical coordinate system ($\theta$, $\phi$) to describe spin-direction, resonant frequency ($\omega$) and spectral linewidth ($\Delta\omega$) can be derived as follows from Landau-Lifshitz-Gilbert equation [1],

$$\omega \approx \sqrt{\det\left[\hat{\Omega}\right]} \approx \sqrt{\omega_{\phi\phi}\omega_{\theta\theta}}, \qquad (S1)$$

$$\Delta\omega \approx \mathrm{Tr}\left[\hat{\Delta}\right] = \Delta\omega_{\phi\phi} + \Delta\omega_{\theta\theta}, \qquad (S2)$$

$$\hat{\Omega} \equiv \begin{pmatrix} \omega_{\phi\phi} & \omega_{\phi\theta} \\ \omega_{\theta\phi} & \omega_{\theta\theta} \end{pmatrix} = \frac{1}{(1+\alpha^2)} \begin{pmatrix} \dfrac{\partial^2 U}{\partial \phi^2} & -\dfrac{1}{S\sin\theta}\dfrac{\partial^2 U}{\partial \theta \partial \phi} \\ -\dfrac{1}{S\sin\theta}\dfrac{\partial^2 U}{\partial \theta \partial \phi} & \dfrac{1}{S\sin\theta}\dfrac{\partial}{\partial \theta}\dfrac{1}{S\sin\theta}\dfrac{\partial U}{\partial \theta} \end{pmatrix}, \quad (S3)$$

$$\hat{\Delta} \equiv \begin{pmatrix} \Delta\omega_{\phi\phi} & \Delta\omega_{\phi\theta} \\ \Delta\omega_{\theta\phi} & \Delta\omega_{\theta\theta} \end{pmatrix} = \frac{\alpha}{S} \begin{pmatrix} \dfrac{\partial}{\partial \phi}\dfrac{1}{\sin^2\theta}\dfrac{\partial U}{\partial \phi} & - \\ - & \dfrac{1}{\sin\theta}\dfrac{\partial}{\partial \theta}\sin\theta\dfrac{\partial U}{\partial \theta} \end{pmatrix}. \quad (S4)$$

Here, $U$, $\alpha$, and $S$ are potential energy, Gilbert damping constant and spin angular momentum, respectively.

For a ferromagnet where potential energy is determined by uniaxial magnetic anisotropy energy ($K$), $\omega \sim K\hbar^{-1}$ can be derived. This can be understood as both $\phi$- and $\theta$- precessional motions are in-phase and are governed only by $K$. Spectral linewidth can be $\Delta\omega_{\phi\phi} + \Delta\omega_{\theta\theta} \sim 2\alpha K\hbar^{-1}$ ($\because \Delta\omega_{\phi\phi} = \Delta\omega_{\theta\theta} \sim \alpha K\hbar^{-1}$). Consider the mode I (optical mode) in a chiral AF metal, where kagome-plane is in $\phi$-plane ($xy$-plane in Fig. 1d in the main text). Here, the resonant frequency can be $\omega_\mathrm{I} \sim \sqrt{JD}\hbar^{-1}$ (see Eq. 3 in the main text for the exact solution). This is because out-of-phase $\phi$-motion ($xy$-plane) is determined by exchange interaction ($J$) and in-phase $\theta$-motion ($z$) is determined by Dzyaloshinskii-Moriya interaction ($D$), where $D$ acts as the magnetic anisotropy for the $\theta$-motion. Spectral linewidth can be $\Delta\omega_{\phi\phi} + \Delta\omega_{\theta\theta} \sim \alpha J\hbar^{-1}$ ($\because \Delta\omega_{\phi\phi} \sim \alpha JS\hbar^{-1}$ and $\Delta\omega_{\theta\theta} \sim \alpha DS\hbar^{-1}$, see Eq. 5 in the main text for the exact solution). For the mode II (collective precession mode) in a chiral AF metal, the resonant frequency can be $\sim \sqrt{KJ}\hbar^{-1}$ (see Eq. 4 in the main text for the exact solution) because in-phase $\phi$-motion and out-of-phase $\theta$-motion are determined by $K$ and $J$, respectively. Spectral linewidth can be $\Delta\omega_{\phi\phi} + \Delta\omega_{\theta\theta} \sim \alpha J\hbar^{-1}$ ($\because \Delta\omega_{\phi\phi} \sim \alpha KS\hbar^{-1}$ and $\Delta\omega_{\theta\theta} \sim \alpha JS\hbar^{-1}$, see Eq. 5 in the main text for the exact solution).

## 2. Sample preparation

Polycrystalline samples were prepared by melting the mixtures of Mn and Sn in an $Al_2O_3$ crucible sealed in an evacuated quartz ampoule in a box furnace at 1050 °C for 6 h. In preparation for single-crystal growth, the obtained polycrystalline materials were crushed into powders, compacted into pellets, and inserted into an $Al_2O_3$ crucible that was subsequently sealed in an evacuated $SiO_2$ ampoule. Single-crystal growth was performed using a single-zone Bridgman furnace with a maximum temperature of 1080 °C and growth speed of 1.5 mm h$^{-1}$. These are exactly the same methods for fabricating the single crystals as those used for the previous study on the Kerr effect [2] and the chiral anomaly due to Weyl fermions [3]. Analysis using inductively coupled plasma spectroscopy showed that the composition of the single crystal is $Mn_{3.07}Sn_{0.93}$. The bulk-$Mn_3Sn$ sample was cut and polished so that the sample had optically smooth surfaces along the $(2\bar{1}\bar{1}0)$ plane. Then the sample was annealed at 600 °C under vacuum (~2×10$^{-6}$ Pa) for 1 h. Without breaking vacuum, 4-nm-AlO$_x$ was subsequently prepared onto the $Mn_3Sn$ surface by electron-beam deposition method to prevent degradation during MOKE measurements.

### 3. Time-resolved magneto-optical Kerr effect (TR-MOKE) measurements

Time-resolved polar MOKE signals were measured in a conventional all-optical pump-probe setup using a Ti-sapphire laser with a regenerated amplifier [4]. The incident light is almost perpendicular to the $Mn_3Sn$ surface, and the polar MOKE signal is proportional to the cluster magnetic octupole component normal to the $Mn_3Sn$ surface. The laser wavelength, pulse width, and repetition rate were 800 nm, 120 fs, 1 kHz, respectively. A penetration depth of the incident light was ~20 nm. The pump beam was modulated with a frequency of 360 Hz using an optical chopper, and the signal was detected using a lock-in amplifier. Spot size and averaged laser power were approximately 1 mW and $\phi$0.66 mm for pump, and 0.05 mW and $\phi$0.14 mm for probe beams, respectively. The delay time dependence of the Kerr rotation signal was recorded. The Kerr rotation signal was detected by the differential method. For the case of the TR-MOKE measurements, non-magnetic backgrounds (e.g. a quadratic background confirmed in the static MOKE results (Fig. 2a in the main text) attributed to an artifact from measurement setup) were cancelled by averaging the data measured for the reversed magnetic field. All measurements were performed at room temperature.

### 4. Characterizing the resonant frequency and the effective damping constant.

To characterize the resonant frequency and the effective damping constant, the obtained signal was fit by an equation describing damped oscillation which is proportional to $\cos(\omega_{I(II)}t+\phi)\exp(-\alpha_{I(II)}\omega_{I(II)}t)$, where $\omega_{I(II)}$, $\phi$ and $\alpha_{I(II)}$ are the resonant frequency, the initial phase, and the effective damping constant of modes I (II), respectively. Because the damped oscillation was confirmed during the relaxation from the ultrafast demagnetization-like behavior, an appropriate background expressing the relaxation process should be subtracted from the raw data. As a function to describe the background, we employed an exponential function, which is depicted in the upper panel of Figs. 3b and 3c in the main text as a black solid curve.

Because of the large effective damping constant ($\alpha_{II}$ = 1.0), the oscillation cannot be recognized at a glance in Fig. 3c in the main text. To increase the accuracy of the fit, a background signal, where there is no spin-wave-induced oscillation, is obtained by measuring a sputter-deposited polycrystalline 50 nm-$Mn_3Sn$ film. Because the *c*-axis of each $Mn_3Sn$ grain is randomly oriented in the film, coherent excitation should not be made. Here, the TR-MOKE experiment is done with the setup and conditions that are different from those employed in the main text using bulk-$Mn_3Sn$. The laser wavelength, pulse width, and repetition rate are 800 nm, 140 fs, 80 MHz, respectively. Pump fluence is ~10 times smaller than that employed in Fig. 3 in the main text. Because of the relatively small Kerr rotation angle for the 800 nm-wavelength (Fig. S1 inset), we have employed a different pulse laser system from the one employed in the main text. We employed magnetron sputtering to prepare a polycrystalline $Mn_3Sn$ thin film and confirmed that the film exhibits anomalous Hall effect and MOKE signal due to the cluster magnetic octupole (Fig. S1 inset). The sample fabrication procedure can be found elsewhere [5]. As shown in Fig. S1, TR-MOKE results indicate the absence of the oscillation signals attributed to mode II. The black solid curve shows a fit using an exponential function. The same background is used for the one in Fig. 3c in the main text.

In Fig. 3c of the main text, there is an oscillation-like signal around 10 ps, which is apparently different from the oscillation signals originating from the modes I and II. As explained above, TR-MOKE results in Fig. S1 indicate the absence of the oscillation signals attributed to modes I and II but a similar oscillation-like signal around 10 ps remains. As this point, we do not have enough information to identify its origin. However, it should be noted that the signal has nothing to do with the coherent spin-wave oscillation.

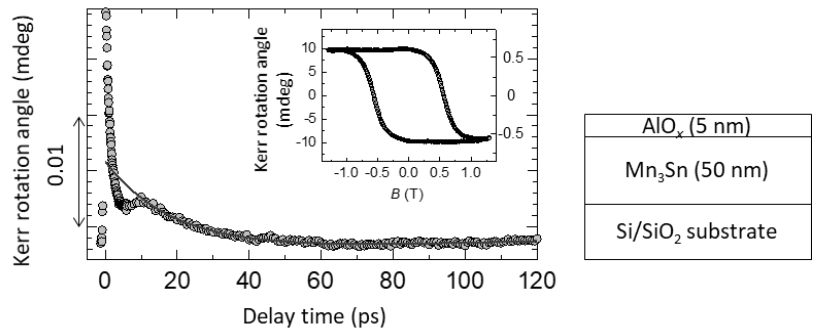

**Figure S1.** TR-MOKE results for a $Mn_3Sn$ thin film. Randomly oriented polycrystalline 50nm-$Mn_3Sn$ is employed for the TR-MOKE measurement. An external magnetic field of ±1.2 T normal to the film plane is applied during the measurements. The black solid curve shows the fit using an exponential function. The background for Fig. 3c in the main text is based on this fit. Inset shows the hysteresis loops of MOKE due to the octupole polarization, where a magnetic field is applied normal to the film plane. The left (right) axis of inset shows the polar Kerr rotation angle measured with 660-nm continuous-wave (800-nm pulse) laser system.

## 5. Coherent spin-wave excitation in Mn$_3$Sn

During the TR-MOKE measurements, an external static magnetic field is applied to direct the octupole polarization along the magnetic field direction. Moreover, the static magnetic field is necessary to adjust the initial phase of each Mn spin to an identical one, which is indispensable for observation of mode I. Figure S2a shows a ground state of the spin structure of Mn$_3$Sn under zero-field while Fig. S2b presents the configuration stabilized under an external magnetic field. As shown in Fig. S2b, an external magnetic field induces the canting of the magnetic moments and aligns polarization of all the octupoles along with the same direction as the initial phase depicted in Fig. 1d in the main text. When the parameters ($J$, $D$, and $K$) are decreased by a pump pulse, the canting angle increases, and the precession can be driven coherently.

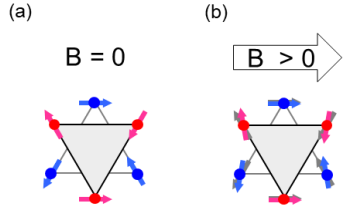

**Figure S2.** (a) Spin structure of Mn$_3$Sn under zero field. (b) Spin structure of Mn$_3$Sn under a static magnetic field.

Here, we consider a model to analyze the octupole oscillations in the TR-MOKE signal after the coherent excitation. As shown in Fig. 2a in the main text, the Kerr rotation angle of the Mn$_3$Sn is negative as a response to the octupole polarization. Therefore, the oscillation should start with increasing the Kerr rotation angle when the phase delay $\phi_0$ is negligible (see the dashed curves in Fig. S3a for $\phi_0 = 0$ deg.). This can be understood as follows. A pump pulse induces a rapid change in the temperature and the magnetic moment, decreasing the anisotropy energy of each magnetic moment determined by the exchange and the spin-orbit interactions. Therefore, the pump pulse increases a spin-canting induced by an external magnetic field (**B**) as explained above. In other words, the pump pulse creates an effective magnetic field parallel to **B**, following the similar mechanism to the case of a ferromagnet [6]. As a result, the torque to drive a coherent precession arises along out-of-kagome-plane direction as indicated in red (Fig. S3a, see below). When the coherent precession starts, the octupole order parameter reduces. For the case of Fig. 3b in the main text, the phase delay ($\phi_0$) cannot be neglected because the transition time for the recovery ($\tau \sim 0.7$ ps) is comparable to the oscillation period ($\sim 1.2$ ps ($\omega_I/2\pi = 0.86$ THz)). Here, let us assume the exponential decay of the torque on each magnetic moment (as shown in Fig. S3b) because this is the simplest and most natural temporal profile of the torque. Then the phase delay can be derived to be $\phi_0 = \tan^{-1}(\omega_I \tau)$ as depicted in Fig. S3c [7]. For the observed recovery time of $\tau = 0.7$ ps, a phase delay is estimated to be around 70-80 deg., which is close to the maximum one ($\phi_0 = 90$ deg.).

Notably, after considering the phase delay, the simulated Kerr rotation angle (solid curve in Fig. S3a) well reproduces the observed oscillation signal in Fig. 3b in the main text.

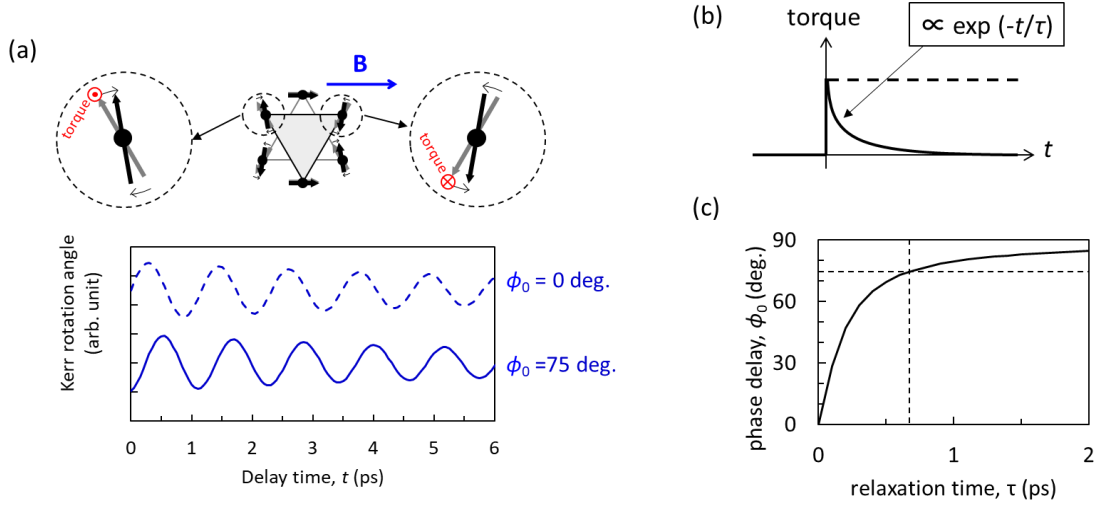

**Figure S3.** (a) Possible MOKE oscillation from the octupole polarization. (b) Temporal profile of a torque used for calculation of Fig. S3c. (c) Calculated phase delay as a function of the relaxation time.

## 6. Theoretical study to determine the spin-wave modes

Spin-wave excitations in Eq. 1 were calculated on the inverse triangular spin structure. First of all, we transform the local frame such that its $z$-axis is parallel to the classical spin direction of the ground state [8]. For the spin operator defined on the new frame ($\mathbf{S'}_{ia} = O_a^{-1} \mathbf{S}_{ia}$), we then perform the Holstein-Primakoff transformation ($\mathbf{S'}_{ia} = M \mathbf{a}_{ia}$), where $O_a$, $M$, and $\mathbf{a}_{ia}$ are respectively given by

$$O_a = \begin{bmatrix} 0 & -\sin\psi_a & \cos\psi_a \\ 0 & \cos\psi_a & \sin\psi_a \\ -1 & 0 & 0 \end{bmatrix}, \quad M = \sqrt{\frac{S}{2}} \begin{bmatrix} 1 & 1 & 0 \\ -i & i & 0 \\ 0 & 0 & \sqrt{\frac{2}{S}} \end{bmatrix}, \quad (S5)$$

and $\mathbf{a}_{ia} = (a_{ia}, a^\dagger_{ia}, S - a^\dagger_{ia} a_{ia})$. Here, the annihilation (creation) operator $a_{ia}$ ($a^\dagger_{ia}$) represents the Holstein-Primakoff boson, and the Hamiltonian (Eq. 1 in the main text) is expressed in terms of $a_{ia}$ and $a^\dagger_{ia}$. Within the linear spin-wave approximation, we neglect higher-order terms of $a_{ia}$ and $a^\dagger_{ia}$, and calculate spin-wave excitation energies keeping them up to the quadratic order.

## 7. Analysis for the torque measurements

Figure S4 shows the schematic diagram of the detailed torque measurements setup of the in-plane magnetic field rotation in the kagome-lattice of Mn$_3$Sn. The setup corresponds to the experiment for Fig. 4b in the main text. Here **B**, **m**, $\phi_B$, and $\phi_m$ show the magnetic field vector, unit vector along with the spontaneous magnetization, magnetic field angle, and angle of the spontaneous magnetization, respectively. Note that the direction of the spontaneous magnetization should be parallel to the octupole polarization. Measurements were conducted at 300 K. Orange curves in Figs. 4a and 4b in the main text show calculated results to reproduce the torque experiments using the following equation:

$$\text{Torque:} \quad -\frac{\partial U}{\partial \phi}\bigg|_{\phi \to \phi_m}, \qquad [\text{S6}]$$

$$U = \frac{K_2}{2}\sin^2\phi + \frac{K_4}{8}\sin^2 2\phi + \frac{K_6}{18}\sin^2 3\phi, \qquad [\text{S7}]$$

where $U$ represents magnetic anisotropy energy. Here, $K_2/2$ ($= 3.4\times10^3$ Jm$^{-3}$), $K_4/8$ ($= -1.8\times10^2$ Jm$^{-3}$), and $K_6/18$ ($= 3.1\times10^2$ Jm$^{-3}$) are the energy barrier heights for two-, four- and six-fold magnetic anisotropy. Therese values are comparable to those obtained in the previous work [9].

From the crystal structure of $D0_{19}$-Mn$_3$Sn, only six-fold magnetic anisotropy should appear. However, we find that in addition to the $K_6$ term, $K_2$ and $K_4$ terms are indispensable to reproduce the experimentally obtained torque data and are strongly field-dependent. Therefore, $K_2$ and $K_4$ terms would be attributed to magnetostriction in Mn$_3$Sn, and an intrinsic magnetic anisotropy at **B** = 0 should be six-fold only, being described by $K_6/18 = 3.1\times10^2$ Jm$^{-3}$. Thus, the energy height ($K_6/18$) determines the thermal stability of the octupole polarization domain at **B** = 0. Because the $K_6$ term corresponds to $S^2K$ in Eq. 1 in the main text, $K$ is calculated to be $3.1\times10^{-4}$ meV per Mn atom with $S = 1.5$.

The saturation magnetization ($M_S$) of Mn$_3$Sn should correspond to the spontaneous magnetization due to spin canting at zero field, determined by $D$ and $J$, and should be distinguished from the magnetization components due to the spin canting induced by the application of an external magnetic field. From the out-of-plane magnetic field rotation measurements (Fig. 4a in the main text), the saturation magnetization can be characterized under the assumption of $M_S B \ll K_2$. From the inset of Fig. 4a in the main text, $M_S$ was estimated be $11.3\times10^{-3}\mu_B$ per f.u. at **B** = 0.

To direct the octupole polarization in a single-domain Mn$_3$Sn from the easy (e.g. $[2\bar{1}\bar{1}0]$) to hard (e.g. $[01\bar{1}0]$) axes, the magnetic field of 3.8 T ($=K_6/M_S$) should be necessary. In a bulk sample, a much smaller magnetic field should be enough to conduct magnetization switching as shown in Fig. 2a inset in the main text because the magnetization switching can be done via a domain wall sweep.

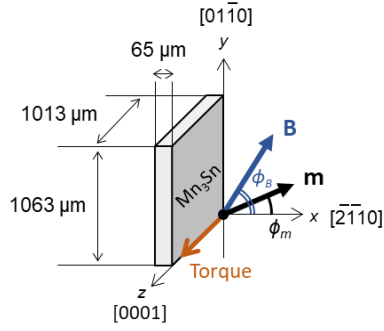

**Figure S4.** Measurement configuration for torque measurements. The schematic diagram of the detailed torque measurements setup of the in-plane magnetic field rotation in the kagome-lattice plane of $Mn_3Sn$.

## 8. Maximum domain wall velocity of the octupole polarization

The maximum domain wall velocity ($v_{max}$) can be derived from Eq. 2 in the main text, where $\alpha = 0$, $K = 0$, $\phi \sim \exp(i\omega t - ikx)$, and $\omega = v_{max}k$ are employed [10].

$$v_{max} = \frac{a_{lat}}{\hbar} S \sqrt{\left(\sqrt{3}D + J\right)\left(\sqrt{3}D + 3J\right)} \qquad [S8]$$

For the case of $Mn_3Sn$, the domain wall propagates with a high velocity without showing the Walker breakdown because of strong magnetic anisotropy due to the Dzyaloshinskii-Moriya interaction, which prevents the out-of-kagome-plane motion (Fig. 4a in the main text) [10].

REFERENCES
1. Suzuki, Y. Spin transfer torque. *Spin current (Oxford University Press)* **2012**, 343.
2. Higo, T.; Man, H.; Gopman, D. B.; Wu, L.; Koretsune, T.; van 't Erve, O. M. J.; Kabanov, Y. P.; Rees, D.; Li, Y.; Suzuki, M.-T.; Patankar, S.; Ikhlas, M.; Chien, C. L.; Arita, R.; Shull, R. D.; Orenstein, J.; Nakatsuji, S. Large mageto-opitical Kerr effect and imaging of magnetic octupole domains in an antiferromagnetic metal. *Nat. Photon.* **2018**, *12*, 73.
3. Kuroda, K.; Tomita, T.; Suzuki, M.-T.; Bareille, C.; Nugroho, A. A.; Goswami, P.; Ochi, M.; Ikhlas, M.; Nakayama, M.; Akebi, S.; Noguchi, R.; Ishii, R.; Inami, N.; Ono, K.; Kumigashira, H.; Varykhalov, A.; Muro, T.; Koretsune, T.; Arita, R.; Shin, S.; Kondo, T.; Nakatsuji, S. Evidence for magnetic Weyl fermions in a correlated metal. *Nat. Mater.* **2017**, *16*, 1090.
4. Iihama, S.; Mizukami, S.; Naganuma, H.; Oogane, M.; Ando, Y.; Miyazaki, T. Gilbert damping constants of Ta/CoFeB/MgO(Ta) thin films measured by optical detection of precessional magnetization dynamics. *Phys. Rev. B* **2014**, *89*, 174416.
5. Higo, T.; Qu, D.; Li, Y.; Chien, C. L.; Otani, Y.; Nakatsuji, S. Anomalous Hall effect in thin films of the Weyl antiferromagnet $Mn_3Sn$. *Appl. Phys. Lett.* **2018**, *113*, 202402.
6. Kirilyuk, A.; Kimel, A. V.; Rasing, T.; Ultrafast optical manipulation of magnetic order.

*Rev. Mod. Phys.* **2010**, *82*, 2731.

7. Schellekens, A. J.; Kuiper, K. C.; de Wit, R. R. J. C.; Koopmans, B. Ultrafast spin-transfer torque driven by femtosecond pulsed-laser excitation. *Nat Commun* **2014**, *5*, 4333.

8. dos Santos, F. J.; dos Santos Dias, M.; Guimarães, F. S. M.; Bouaziz, J.; Lounis, S.; Spin-resolved inelastic electron scattering by spin waves in noncollinear magnets. *Phys. Rev. B* **2018**, *97*, 024431.

9. Duan, T. F.; Ren, W. J.; Liu, W. L.; Li, S. J.; Liu, W.; Zhang, Z. D. Magnetic anisotropy of single-crystalline $Mn_3Sn$ in triangular and helix-phase states. *Appl. Phys. Lett.* **2014**, *107*, 082403.

10. Nomoto, T.; Arita, R. Cluster multipole dynamics in noncollinear antiferromagnets. *Phys. Rev. Research* **2020**, *2*, 012045(R).